\documentclass[12pt]{article}          
\usepackage{subcaption}
\usepackage{graphicx}
\usepackage[export]{adjustbox}
\usepackage[margin=1in]{geometry}    
\usepackage[ruled,vlined]{algorithm2e}
\usepackage{multirow}
\usepackage[noend]{algpseudocode}
\usepackage{subcaption}
\usepackage{amsmath,amssymb,amsfonts}
\usepackage{xcolor}
\usepackage{hyperref}
\usepackage{comment}
\usepackage{multicol}
\usepackage{xcolor}
\usepackage{MnSymbol}
\usepackage{float}
\usepackage{tikz}
\usepackage{booktabs}
\usepackage{longtable}
\usepackage{forest}
\useforestlibrary{edges}
\usepackage{enumitem}

\begin{document}
	
	\title{BRISC: Annotated Dataset for Brain Tumor Segmentation and Classification}

\author{
    Amirreza Fateh\textsuperscript{1}, 
    Yasin Rezvani\textsuperscript{2}, 
    Sara Moayedi\textsuperscript{2}, 
    Sadjad Rezvani\textsuperscript{2}, \\
    Fatemeh Fateh\textsuperscript{3}, 
    Mansoor Fateh\textsuperscript{2,*}, 
    Vahid Abolghasemi\textsuperscript{4,*}
}

	
	
	\maketitle

\begin{center}
    \small
    \textsuperscript{1}School of Computer Engineering, Iran University of Science and Technology (IUST), Tehran, Iran\\
    \textsuperscript{2}Faculty of Computer Engineering, Shahrood University of Technology, Shahrood, Iran\\
    \textsuperscript{3}Northern Care Alliance NHS Foundation Trust (NCA), Manchester, UK\\
    \textsuperscript{4}School of Computer Science and Electronic Engineering, University of Essex, Colchester, UK\\
    
    \vspace{0.3cm}
    \texttt{amirreza\_fateh@comp.iust.ac.ir}\\
    \texttt{yasinrezvani@shahroodut.ac.ir}\\
    \texttt{sara\_moayedi@shahroodut.ac.ir}\\
    \texttt{sadjadrezvani@shahroodut.ac.ir}\\
    \texttt{fatemeh.fateh@nca.nhs.uk}\\
    \texttt{mansoor\_fateh@shahroodut.ac.ir}\\
    \texttt{v.abolghasemi@essex.ac.uk}\\
    
    \vspace{0.2cm}
    \textsuperscript{*}Corresponding authors
\end{center}
    
	\begin{abstract}
		\small
		\color{black}
        Accurate segmentation and classification of brain tumors from Magnetic Resonance Imaging (MRI) remain key challenges in medical image analysis, primarily due to the lack of high-quality, balanced, and diverse datasets with expert annotations. In this work, we address this gap by introducing BRISC, a dataset designed for brain tumor segmentation and classification tasks, featuring high-resolution segmentation masks. The dataset comprises 6,000 contrast-enhanced T1-weighted MRI scans, which were collated from multiple public datasets that lacked segmentation labels. Our primary contribution is the subsequent expert annotation of these images, performed by certified radiologists and physicians. It includes three major tumor types, namely glioma, meningioma, and pituitary, as well as non-tumorous cases. Each sample includes high-resolution labels and is categorized across axial, sagittal, and coronal imaging planes to facilitate robust model development and cross-view generalization. To demonstrate the utility of the dataset, we provide benchmark results for both tasks using standard deep learning models. The BRISC dataset is made publicly available. \color{black} datasetlink: \url{https://www.kaggle.com/datasets/briscdataset/brisc2025/}

	\end{abstract}

\section*{Background \& Summary} \label{background}

Brain tumors are among the most critical medical conditions, requiring accurate and timely diagnosis for effective treatment and management \cite{ge2024open,usman2024brain}. Magnetic Resonance Imaging (MRI) plays a critical tool in diagnosing and monitoring brain tumors, owing to its non-invasive imaging capabilities and ability to provide detailed visualization of brain structures \cite{sun2025low,zhu20257,gong2024multi}. Despite significant advancements in medical imaging technologies, developing automated systems for tumor detection and segmentation remains a major challenge \cite{dorosti2025high,li2025ddevenet,rezvani2025fusionlungnet}. This challenge arises primarily from the scarcity of high-quality labeled datasets designed for these tasks. Additionally, the complexity and variability of tumor appearances across patients further complicate accurate segmentation and classification \cite{zhang2024application,askari2025enhancing,FATEH2025105672}.

Existing brain tumor segmentation datasets, such as the Brain Tumor Segmentation (BraTS) \cite{menze2014multimodal}, \color{black}Cheng \cite{Cheng2017} brain tumor dataset\color{black}, and others, have significantly advanced the development of automated segmentation models. However, several limitations in these datasets drive the need for novel datasets to address emerging challenges in the field. For example, the BraTS dataset is widely used and comprehensive. However, it depends on pre-processed, standardized data that may not reflect the real-world variability in MRI acquisition protocols across institutions. Additionally, BraTS primarily focuses on gliomas and lacks representation of other tumor types, potentially limiting the generalizability of models trained on it \cite{ghaffari2019automated,labella2024analysis}. \color{black}The Cheng \cite{Cheng2017} brain tumor dataset\color{black}, on the other hand, suffers from class imbalance and limited diversity in imaging conditions and patient demographics, which can restrict model robustness \cite{de2024enhanced}. Many publicly available datasets face issues with inconsistent labeling. These inconsistencies can negatively affect the training and evaluation of segmentation models \cite{sarkar2024evolution,khajeha2025advancing,fakhima2025covsgnet,rezvani2024single}. These limitations underscore the necessity of introducing a new dataset that offers balanced class distributions, multi-institutional diversity, and high-quality expert annotations to enhance the reliability and generalizability of automated brain tumor segmentation models. Additionally, including class labels for classification tasks, such as identifying glioma, meningioma, pituitary, and non-tumorous cases, enhances the dataset’s utility and supports broader real-world applications in brain tumor analysis.

To address these gaps, we present the BRISC dataset: a large-scale, balanced, and expert-annotated MRI dataset designed for both segmentation and classification of brain tumors. BRISC includes 6,000 contrast-enhanced T1-weighted MRI scans across four categories—glioma, meningioma, pituitary, and non-tumorous cases—covering multiple anatomical planes (axial, coronal, sagittal). The dataset emphasizes consistent quality, balanced distributions, and multi-institutional diversity, making it suitable for developing models that generalize across clinical settings.

\begin{table}[h]
	\centering
	\caption{Class distribution in the training and testing parts of BRISC}
	\renewcommand{\arraystretch}{1.5}
	\begin{tabular}{|c|c|c|c|}
		\hline
		\textbf{Class}      & \textbf{Training Images} & \textbf{Testing Images} & \textbf{Total Images} \\ \hline
		Glioma              & 1,147                   & 254                     & 1,401                 \\ \hline
		Meningioma          & 1,329                   & 306                     & 1,635                 \\ \hline
		Pituitary           & 1,457                   & 300                     & 1,757                 \\ \hline
		non-tumorous            & 1,067                   & 140                     & 1,207                 \\ \hline
		\textbf{Total}      & \textbf{5,000}          & \textbf{1,000}          & \textbf{6,000}        \\ \hline
	\end{tabular}
	\label{tab:distribution}
\end{table}

\begin{table}[h]
	\centering
	\caption{Class distribution based on MRI planes in the training and testing parts of BRISC}
	\renewcommand{\arraystretch}{1.5} 
	\begin{tabular}{|c|ccc|ccc|}
		\hline
		\multirow{2}{*}{Class\textbackslash{}Plane} & \multicolumn{3}{c|}{\textbf{Train}} & \multicolumn{3}{c|}{\textbf{Test}} \\ \cline{2-7} 
		& \multicolumn{1}{c|}{Axial} & \multicolumn{1}{c|}{Coronal} & Sagittal & \multicolumn{1}{c|}{Axial} & \multicolumn{1}{c|}{Coronal} & Sagittal \\ \hline
		Glioma & \multicolumn{1}{c|}{347} & \multicolumn{1}{c|}{428} & 372 & \multicolumn{1}{c|}{85} & \multicolumn{1}{c|}{81} & 88 \\ \hline
		Meningioma & \multicolumn{1}{c|}{423} & \multicolumn{1}{c|}{426} & 480 & \multicolumn{1}{c|}{134} & \multicolumn{1}{c|}{89} & 83 \\ \hline
		Pituitary & \multicolumn{1}{c|}{428} & \multicolumn{1}{c|}{496} & 533 & \multicolumn{1}{c|}{116} & \multicolumn{1}{c|}{98} & 86 \\ \hline
		non-tumorous & \multicolumn{1}{c|}{352} & \multicolumn{1}{c|}{310} & 405 & \multicolumn{1}{c|}{52} & \multicolumn{1}{c|}{48} & 40 \\ \hline
		\begin{tabular}[c]{@{}c@{}}Total per\\ plane\end{tabular} & \multicolumn{1}{c|}{1550} & \multicolumn{1}{c|}{1660} & 1790 & \multicolumn{1}{c|}{387} & \multicolumn{1}{c|}{316} & 297 \\ \hline
		\textbf{Total} & \multicolumn{3}{c|}{\textbf{5000}} & \multicolumn{3}{c|}{\textbf{1000}} \\ \hline
	\end{tabular}
	\label{tab:planes_distribution}
\end{table}

\section*{Methods}

The BRain tumor Image Segmentation and Classification (BRISC) dataset has been meticulously curated to address key challenges in brain tumor research, particularly in the domains of segmentation and classification tasks. It provides a balanced, high-quality collection of MRI data, annotated for both research and clinical applications. The dataset includes images with labels for four categories: Glioma, Meningioma, Pituitary tumors, and non-tumorous. By focusing on comprehensive data collection and rigorous annotation processes, the dataset aims to advance the development of robust machine learning models in medical imaging.

\subsection*{Motivation and Aims}

The primary goal of collecting and releasing this dataset is to overcome limitations observed in existing brain tumor datasets, such as class imbalance, lack of diversity, and annotation inconsistencies. While datasets like BraTS have driven significant advancements in glioma segmentation, their exclusive focus on gliomas and reliance on pre-processed data limit their generalizability to other tumor types and real-world scenarios. Our dataset expands the scope by incorporating multiple tumor types and includes a "non-tumorous" class to aid in broader diagnostic tasks. This addition makes the dataset highly versatile, enabling its use in applications ranging from multi-class tumor classification to binary tumor detection.

\subsection*{Dataset Composition and Planar Distributions}
The dataset comprises 6,000 MRI images, divided into training and testing sets, as detailed in Table \ref{tab:distribution}. This structured division ensures robust evaluation metrics while providing ample data for training advanced machine learning models. For the training dataset, the total number of images across the planes is 5,000, and for testing, the total is 1,000.

In addition to the class-based distribution, we provide another form of distribution, which is dataset composition by MRI planes. This breakdown categorizes images into Coronal, Sagittal, and Axial planes, helping to analyze how different orientations are represented in the dataset. As shown in Table \ref{tab:planes_distribution}, the distribution of different MRI planes is nearly uniform, similar to the distribution of different classes in Table \ref{tab:distribution}. This balanced distribution ensures that no particular class or plane is overrepresented, which is crucial for preventing model bias and improving generalization.

\subsection*{Data Preprocessing}
\color{black}The original collection contains 7,023 brain MR images across four classes: glioma, meningioma, pituitary tumour, and non-tumorous. We then applied the following steps to ensure consistency and quality:
\color{black}

\color{black}
\begin{description}[font=$\bullet$\scshape\bfseries]
	\item \textbf{Sequence harmonization}: Only T1-weighted MRI sequences were retained to ensure consistency.
    \item \textbf{Label and mask verification}: A radiologist and a physician reviewed images to identify incorrect or inconsistent labels; images with tumour labels unsupported by the visible image, or with empty/misaligned masks, were removed.
    \item \textbf{Artefact/corruption screening}: Corrupted files and images with severe artefacts were excluded.
    \item \textbf{De-duplication and redundancy control}: Exact duplicates and near-duplicates (including consecutive images from very short series) were removed; de-duplication was completed before any train/test split to avoid leakage and over-representation.
    \item \textbf{Standardization}: Images were resized and borders/margins adjusted for consistent spatial dimensions.
\end{description}
\color{black}

\subsection*{Imaging Details}
All images in the dataset are T1-weighted contrast-enhanced MRI scans, selected specifically from the "Brain Tumor MRI Dataset" (Kaggle) \cite{nickparvar2021brain}. Although the original dataset included some T2-weighted images, we exclusively selected T1-weighted scans for their superior ability to highlight tumor boundaries effectively. Another notable characteristic of this dataset is the length of MRI sequences. While typical brain MRI studies often consist of longer sequences, the majority of sequences in this dataset were notably short, ranging from 1 to 5 images per sequence. Sequences with only one image were excluded, as even experienced radiologists and physicians found it challenging to identify tumors accurately in these cases.

\color{black}
The original dataset provided only a train/test split and did not include any information about patients, sequences, or slices, making it impossible to definitively link images to the same subject. To minimize the risk of the same person appearing in both train and test sets, our annotators manually reviewed images as much as possible—through visual similarity checks and metadata cross-referencing—to separate likely same-subject cases. While complete subject-level independence cannot be guaranteed due to the source limitations, this conservative approach ensures no obvious same-patient images cross splits; multiple images from the same subject may therefore be present within a single split. This is documented in the repository metadata.\color{black}

\subsection*{Annotation Process}

\color{black}
The dataset underwent a meticulous annotation and review process to ensure accuracy and reliability. Annotation was performed using the AnyLabeling tool \cite{nguyen2024anylabeling}, which facilitated precise delineation of tumorous and non-tumorous regions. Each image was reviewed and refined multiple times with input from a certified physician and radiologist. Key steps in the annotation process included:

\begin{description}[font=$\bullet$\scshape\bfseries]
    \item \textbf{Tumor Mask Refinement}: Using AnyLabeling \cite{nguyen2024anylabeling}, regions corresponding to tumorous lesions were iteratively refined to ensure accurate segmentation masks.
	\item \textbf{Class Verification}: The “non-tumorous” class was reviewed in detail. It includes both completely healthy brains and scans with non-neoplastic, space-occupying lesions (e.g., abscesses, cysts). Images misclassified in the original datasets were corrected or removed as needed.
    \item \textbf{Consensus Reviews}: Annotation was carried out by a trained team and verified under the supervision of a certified radiologist and physician. Discrepancies were detected through visual comparison and overlap inspection in AnyLabeling and resolved collaboratively during review sessions. A quality assessment on a representative subset showed a mean Dice coefficient of 0.924 between initial and expert‑verified masks. Approximately 4.8\% of images required correction, after which full consensus and final approval were achieved for all BRISC annotations.
\end{description}

\color{black}

\begin{figure}[h]
	\centering
	\resizebox{0.9\textwidth}{!}{%
		\begin{forest}
			for tree={
				font=\small\ttfamily,
				grow'=0,
				child anchor=west,
				parent anchor=south,
				anchor=west,
				calign=first,
				l=0pt,
				s sep=0pt,
				edge path={
					\noexpand\path [draw, \forestoption{edge}]
					(!u.south west) +(5pt,0) |- (.child anchor) \forestoption{edge label};
				},
				before typesetting nodes={
					if n=1
					{insert before={[,phantom]}}
					{}
				},
				fit=band,
			}
			[/brisc2025/
			[classification\_task/
			[train/
			[glioma/]
			[meningioma/]
			[no\_tumor/]
			[pituitary/]
			]
			[test/
			[glioma/]
			[meningioma/]
			[no\_tumor/]
			[pituitary/]
			]
			]
			[segmentation\_task/
			[train/
			[images/]
			[masks/]
			]
			[test/
			[images/]
			[masks/]
			]
			]
			]
		\end{forest}
	}
	\caption{Directory structure of the BRISC dataset.}
	\label{fig:directory_structure}
\end{figure}

\color{black}
\subsection*{Limitations and Intended Use}
BRISC comprises only contrast‑enhanced T1‑weighted MRI images collected from multiple public datasets that lack comprehensive acquisition metadata such as scanner type, field strength, or sequence parameters. Consequently, detailed harmonization across hardware and protocol variations was not achievable. The dataset is intended primarily for research in algorithm benchmarking, model comparison, and methodological development for brain‑tumour segmentation and classification tasks. It is not designed or validated for direct clinical diagnostic use. Users should be aware that models trained on BRISC may experience domain‑shift when applied to non‑contrast T1‑weighted, T2‑weighted, or institution‑specific datasets, and additional data normalization or fine‑tuning is recommended to mitigate such effects.

\section*{Data Records}

The BRISC (BRain tumor Image Segmentation and Classification) dataset is available in Kaggle \cite{fateh2025brisc}. The dataset release includes:
\begin{itemize}[leftmargin=*,noitemsep]
    \item Dataset images
    \item Manifest file (\texttt{manifest.csv})
    \item JSON metadata (\texttt{manifest.json})
    \item File checksums and per-file metadata
\end{itemize}

The public release follows the directory structure shown in Figure~\ref{fig:directory_structure}.

\begin{figure}[]
	\centering
	\renewcommand{\arraystretch}{1.5}
	\resizebox{0.6\textwidth}{!}{
		\includegraphics[width=\textwidth]{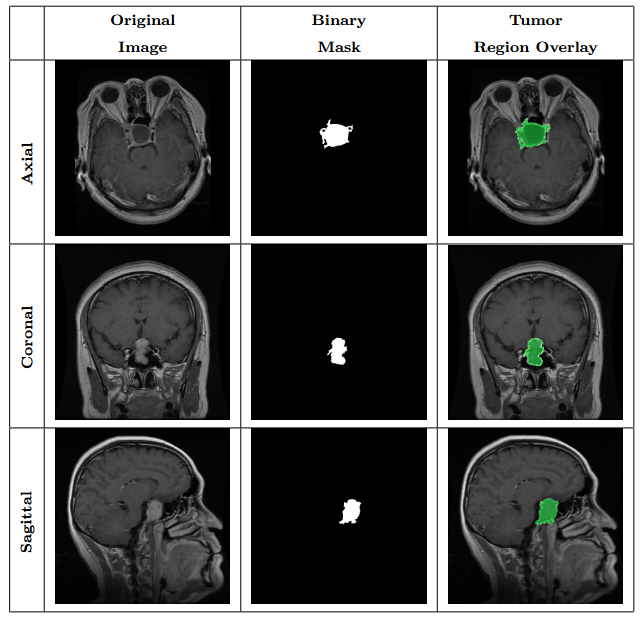}
	}
	\caption{Samples of Pituitary segmentation across different imaging planes}
	\label{tab:pituitary_seg}
\end{figure}

\subsection*{File Naming Convention}
Filenames follow this pattern:\\[0.5em]
\noindent\texttt{brisc2025\_<split>\_<index>\_<tumor\_code>\_}\\
\texttt{<plane\_code>\_<sequence>.<ext>}\\[0.5em]
\noindent\textbf{Example:} \texttt{brisc2025\_test\_00010\_gl\_ax\_t1.jpg}\\
For segmentation pairs, images and masks share the same basename (image: \texttt{.jpg}, mask: \texttt{.png}).

\subsection*{Metadata Fields}
Each image row in the manifest includes:
\begin{itemize}[leftmargin=*,noitemsep]
    \item \texttt{relative\_path}, \texttt{filename}, \texttt{task}, \texttt{split}
    \item \texttt{index}, \texttt{tumor\_code}, \texttt{tumor\_label}
    \item \texttt{plane\_code}, \texttt{plane\_label} (ax=axial, co=coronal, sa=sagittal)
    \item \texttt{sequence} (e.g., t1=T1-weighted)
    \item Spatial dimensions (\texttt{width}, \texttt{height})
    \item \texttt{file\_size\_bytes}, \texttt{sha256} checksum
\end{itemize}

Mask rows have \texttt{is\_mask = True} and include \texttt{linked\_image} pointing to the matched image.

BRISC is derived from Nickparvar ``Brain Tumor MRI Dataset''  \cite{nickparvar2021brain}, which aggregates:
\begin{itemize}[leftmargin=*,noitemsep]
	\item Cheng brain tumor dataset \cite{Cheng2017}
	\item SARTAJ \cite{bhuvaji2021brain}
	\item Br35H \cite{hamada2020br35h}
\end{itemize}
Selection and exclusion rules applied during curation are documented in the Methods section.

\section*{Data Overview}
BRISC contains 2D single-slice T1-weighted brain MRI images stored as JPEG files, with corresponding pixel‑wise segmentation masks stored as PNG files. The dataset is organized into two separate tasks:

\begin{description}
	\item[\textbf{1) Classification task:}]
	Balanced image‑level classification dataset for four diagnostic classes: glioma, meningioma, pituitary tumor, and no\_tumor. Contains 6,000 JPEG images (5,000 training, 1,000 test).
	
	\item[\textbf{2) Segmentation task:}]
	Pixel‑wise tumor annotation task. Contains 4,793 image 	files with exact paired masks. Figure~\ref{tab:pituitary_seg} shows an example image from the dataset together with its mask.
\end{description}

\color{black}

\section*{Technical Validation}
\color{black}
Establishing baseline performance is a critical step in evaluating any newly proposed dataset, as it sets a reference point for further research and model development \cite{cheng2024interactive}. To validate the effectiveness and versatility of the BRISC dataset, we conducted experimental evaluations on segmentation and classification tasks.

This section presents a comprehensive performance analysis of several standard baseline models. We report their performance metrics to establish clear benchmarks for the dataset. These results demonstrate the dataset's quality and utility for developing and testing new models in medical imaging.
\color{black}
\subsection*{Evaluation Metrics}

\subsubsection*{Segmentation Metric}
\color{black}
In this part, we detail the evaluation metrics employed to assess the performance of segmentation models on the BRISC dataset. These metrics provide comprehensive insights into model efficacy for these two distinct tasks.\color{black}

\paragraph{\textbf{Intersection over Union (IoU)}}

Intersection over Union (IoU), also known as the Jaccard Index, is a fundamental metric for evaluating binary segmentation tasks. It quantifies the overlap between the predicted tumor regions and the ground truth, normalized by their union~\cite{rezatofighi2019generalized}. For binary segmentation, IoU is computed as shown in Equation \ref{eq:iou}.

\begin{equation}
	\text{IoU} = \frac{\sum_{i=1}^{N} y_i \hat{y}_i}{\sum_{i=1}^{N} y_i + \sum_{i=1}^{N} \hat{y}_i - \sum_{i=1}^{N} y_i \hat{y}_i + \epsilon},
	\label{eq:iou}
\end{equation}

where \( y_i \in \{0,1\} \) denotes the ground truth label, \( \hat{y}_i \in \{0,1\} \) represents the predicted label (after thresholding), and \( \epsilon \) is a small constant added for numerical stability.

As shown in Equation~\ref{eq:iou}, this formulation captures the pixel-wise overlap between the predicted and actual tumor regions and is particularly effective for evaluating segmentation quality, especially along object boundaries.

\subsubsection*{Classification Metrics}

As commonly employed in the evaluation of multi-class classification models, metrics such as Accuracy, Precision, Recall, and F1-Score are widely utilized due to their effectiveness in assessing performance across diverse tasks \cite{goodfellow2016deep,he2016deep,dosovitskiy2020image}. These metrics are defined below to ensure a comprehensive understanding of their applicability to our dataset.

\paragraph{\textbf{Accuracy}}

Accuracy measures the overall correctness of predictions across all four classes. It is defined as:

\begin{equation}
	\text{Accuracy} = \frac{\sum_{i=1}^{C} \text{Correct Predictions for Class } i}{\text{Total Samples}}
\end{equation}

where \(C\) denotes the total number of classes, and "Correct Predictions for Class \(i\)" represents the samples correctly classified as class \(i\).

\paragraph{\textbf{Precision}}

Precision quantifies the proportion of correctly predicted positive instances for each class. For class \(i\), Precision is defined as:

\begin{equation}
	\text{Precision}_i = \frac{\text{TP}_i}{\text{TP}_i + \text{FP}_i}
\end{equation}

In multi-class classification, Precision is averaged using either macro-averaging or weighted-averaging:

\begin{equation}
	\text{Macro Precision} = \frac{1}{C} \sum_{i=1}^{C} \text{Precision}_i
\end{equation}

\begin{equation}
	\text{Weighted Precision} = \frac{\sum_{i=1}^{C} w_i \cdot \text{Precision}_i}{\sum_{i=1}^{C} w_i}
\end{equation}

where \(w_i\) represents the proportion of samples in class \(i\).

\paragraph{\textbf{Recall}}

Recall, or Sensitivity, measures the proportion of actual positive instances correctly identified by the model. For class \(i\), Recall is defined as:

\begin{equation}
	\text{Recall}_i = \frac{\text{TP}_i}{\text{TP}_i + \text{FN}_i}
\end{equation}

For multi-class classification, Recall is averaged similarly to Precision:

\begin{equation}
	\text{Macro Recall} = \frac{1}{C} \sum_{i=1}^{C} \text{Recall}_i
\end{equation}

\begin{equation}
	\text{Weighted Recall} = \frac{\sum_{i=1}^{C} w_i \cdot \text{Recall}_i}{\sum_{i=1}^{C} w_i}
\end{equation}

\paragraph{\textbf{F1-Score}}

The F1-Score is the harmonic mean of Precision and Recall. For class \(i\), it is defined as:

\begin{equation}
	\text{F1-Score}_i = 2 \cdot \frac{\text{Precision}_i \cdot \text{Recall}_i}{\text{Precision}_i + \text{Recall}_i}
\end{equation}

For multi-class classification, F1-Score is averaged as follows:

\begin{equation}
	\text{Macro F1-Score} = \frac{1}{C} \sum_{i=1}^{C} \text{F1-Score}_i
\end{equation}

\begin{equation}
	\text{Weighted F1-Score} = \frac{\sum_{i=1}^{C} w_i \cdot \text{F1-Score}_i}{\sum_{i=1}^{C} w_i}
\end{equation}

By calculating these metrics per class and aggregating them through macro- or weighted-averaging, we ensure a detailed evaluation of model performance, particularly in datasets with imbalanced class distributions.

\subsection*{Comparison}

\begin{table}[h]
	\centering
	\caption{IoU (\%) for Brain Tumor Segmentation Models on Different Tumor Types. Weighted mIoU is calculated as a weighted average based on the number of samples per tumor type: Glioma, Meningioma, Pituitary.}
	\renewcommand{\arraystretch}{1.5}
	\begin{tabular}{lccc|c}
		\hline
		\textbf{Model} & \begin{tabular}[c]{@{}c@{}}\textbf{mIoU} \\ \textbf{Glioma}\end{tabular} & \begin{tabular}[c]{@{}c@{}}\textbf{mIoU} \\ \textbf{Meningioma}\end{tabular} & \begin{tabular}[c]{@{}c@{}}\textbf{mIoU} \\ \textbf{Pituitary}\end{tabular} & \begin{tabular}[c]{@{}c@{}}\textbf{Weighted} \\ \textbf{mIoU}\end{tabular} \\ \hline
		UNet~\cite{ronneberger2015u}            & 69.7                  & 77.1                      & 79.3                     & 75.7                      \\
		UNet++~\cite{zhou2018unet++}            & 71.7                  & 74.2                      & 79.7                     & 75.3                      \\
		MANet~\cite{fan2020ma}            & 72.4                  & 77.5                      & 78.0                     & 76.2                      \\
		LinkNet~\cite{chaurasia2017linknet}            & 71.7                  & 74.8                      & 79.0                     & 75.3                      \\
		DeepLabV3+~\cite{chen2018encoder}            & 72.0                 & 77.5                      & 78.7                     & 76.3                      \\
		PAN~\cite{li2018pyramid}            & 72.0                  & 74.5                      & 80.7                     & 75.9                      \\
		EINet~\cite{li2022einet}            & 73.6                  & 78.4                      & 80.3                     & 77.7                      \\
		EU-Net~\cite{patel2021enhanced}            & 71.7                  & 76.1                      & 78.3                     & 75.6                      \\
		DAD~\cite{li2022towards}            & 75.2                  & 80.4                      & 82.3                     & 79.5                      \\
		BASNet~\cite{qin2021boundary}            & 74.0                  & 77.5                      & 81.7                     & 77.9                      \\
		SaberNet~\cite{saber2024efficient}    & 74.0                  & 82.4                      & 84.3                     & 80.6                      \\
		ABANet~\cite{rezvani2024abanet} & 72.4                  & 80.4                      & 84.7                     & 79.5                     \\
		\hline
	\end{tabular}
	\label{tab:iou_segmentation}
\end{table}

\subsubsection*{Segmentation results}

\color{black}
To establish segmentation benchmarks for the dataset, we conducted a comparative study against a diverse set of brain tumor segmentation models, including traditional convolutional architectures, attention-enhanced methods, and transformer-based approaches. The baselines include UNet~\cite{ronneberger2015u}, UNet++~\cite{zhou2018unet++}, LinkNet~\cite{chaurasia2017linknet}, MANet~\cite{fan2020ma}, DeepLabV3+~\cite{chen2018encoder}, PAN~\cite{li2018pyramid}, EINet~\cite{li2022einet}, EU-Net~\cite{patel2021enhanced}, DAD~\cite{li2022towards}, and BASNet~\cite{qin2021boundary}, as well as two recent transformer-enhanced models, SaberNet~\cite{saber2024efficient} and ABANet~\cite{rezvani2024abanet}.

Each model was evaluated using the mean Intersection over Union (mIoU) metric for three tumor types: \textit{Glioma}, \textit{Meningioma}, and \textit{Pituitary}. Furthermore, we report a \textit{weighted mIoU}, which is calculated based on the proportion of samples belonging to each tumor type, providing a more representative performance indicator across the dataset.

As summarized in Table~\ref{tab:iou_segmentation}, the results establish a strong set of baselines. Transformer-based models, such SaberNet, achieved the highest scores, with weighted mIoUs of 80.6\% respectively. This suggests that architectures adept at capturing multi-scale contextual features perform well on this dataset. More traditional architectures like UNet provided a solid baseline with a weighted mIoU of 75.7\%.

The reported weighted mIoU is calculated as a weighted average based on the number of samples in each tumor class to provide a more realistic assessment under dataset imbalance. Unlike simple arithmetic means, the weighted mean better reflects the overall segmentation performance in real-world clinical distributions.

It is important to emphasize that the primary goal of this work is to introduce and validate a new brain tumor segmentation dataset, which is designed to support the development of robust and generalizable medical segmentation models. These results serve as a foundational step, and future research is expected to build upon this dataset to explore a broader range of models, training protocols, and evaluation settings.

\color{black}

\begin{table}[]
	\centering
	\caption{Per-Class and Average Classification Performance (\%) for Brain Tumor Classification Models. Metrics are reported as mean $\pm$ standard deviation over three runs.}
	
	{\tiny
		\resizebox{0.77\textwidth}{!}{
			\begin{tabular}{llrrrr}
				\toprule
				Model & Class & Precision & Recall & F1-Score & Accuracy  \\
				\midrule
				ResNet50 & glioma & 0.9868 $\pm$ 0.0098 & 0.9751 $\pm$ 0.0164 & 0.9808 $\pm$ 0.0103 & - \\
				& meningioma & 0.9815 $\pm$ 0.0150 & 0.9673 $\pm$ 0.0226 & 0.9741 $\pm$ 0.0087 & - \\
				& no\_tumor & 0.9906 $\pm$ 0.0107 & 0.9952 $\pm$ 0.0083 & 0.9929 $\pm$ 0.0035 & - \\
				& pituitary & 0.9756 $\pm$ 0.0285 & 0.9967 $\pm$ 0.0000 & 0.9859 $\pm$ 0.0146 & - \\
				& Macro Avg & 0.9836 $\pm$ 0.0064 & 0.9836 $\pm$ 0.0077 & 0.9834 $\pm$ 0.0072 & - \\
				& Weighted Avg & 0.9823 $\pm$ 0.0076 & 0.9820 $\pm$ 0.0080 & 0.9820 $\pm$ 0.0080 & 0.9820 $\pm$ 0.0080 \\
				\midrule
				ResNet101 & glioma & 0.9726 $\pm$ 0.0347 & 0.9869 $\pm$ 0.0082 & 0.9794 $\pm$ 0.0138 & - \\
				& meningioma & 0.9879 $\pm$ 0.0081 & 0.9575 $\pm$ 0.0279 & 0.9722 $\pm$ 0.0106 & - \\
				& no\_tumor & 0.9883 $\pm$ 0.0107 & 0.9905 $\pm$ 0.0165 & 0.9893 $\pm$ 0.0037 & - \\
				& pituitary & 0.9793 $\pm$ 0.0113 & 0.9956 $\pm$ 0.0020 & 0.9874 $\pm$ 0.0066 & - \\
				& Macro Avg & 0.9820 $\pm$ 0.0074 & 0.9826 $\pm$ 0.0093 & 0.9821 $\pm$ 0.0086 & - \\
				& Weighted Avg & 0.9815 $\pm$ 0.0085 & 0.9810 $\pm$ 0.0092 & 0.9810 $\pm$ 0.0092 & 0.9809 $\pm$ 0.0092 \\
				\midrule
				DenseNet121 & glioma & 0.4838 $\pm$ 0.4753 & 0.5879 $\pm$ 0.5095 & 0.5197 $\pm$ 0.4731 & - \\
				& meningioma & 0.6640 $\pm$ 0.5751 & 0.2800 $\pm$ 0.4569 & 0.3178 $\pm$ 0.4966 & - \\
				& no\_tumor & 0.5324 $\pm$ 0.4095 & 0.6976 $\pm$ 0.4871 & 0.4721 $\pm$ 0.4204 & - \\
				& pituitary & 0.4502 $\pm$ 0.4128 & 0.6422 $\pm$ 0.5574 & 0.5259 $\pm$ 0.4678 & -\\
				& Macro Avg & 0.5326 $\pm$ 0.4550 & 0.5519 $\pm$ 0.3376 & 0.4589 $\pm$ 0.4310 & - \\
				& Weighted Avg & 0.5356 $\pm$ 0.4650 & 0.5253 $\pm$ 0.3850 & 0.5253 $\pm$ 0.3850 & 0.4531 $\pm$ 0.4390 \\
				\midrule
				DenseNet169 & glioma & 0.9543 $\pm$ 0.0690 & 0.3543 $\pm$ 0.5356 & 0.3840 $\pm$ 0.5173 & - \\
				& meningioma & 0.7522 $\pm$ 0.2090 & 0.8007 $\pm$ 0.2607 & 0.7560 $\pm$ 0.2021 & - \\
				& no\_tumor & 0.4841 $\pm$ 0.4507 & 0.9738 $\pm$ 0.0393 & 0.5754 $\pm$ 0.3763 & - \\
				& pituitary & 0.3333 $\pm$ 0.5774 & 0.3322 $\pm$ 0.5754 & 0.3328 $\pm$ 0.5764 & - \\
				& Macro Avg & 0.6310 $\pm$ 0.3116 & 0.6152 $\pm$ 0.3305 & 0.5121 $\pm$ 0.4160 & - \\
				& Weighted Avg & 0.6404 $\pm$ 0.3020 & 0.5710 $\pm$ 0.3689 & 0.5710 $\pm$ 0.3689 & 0.5093 $\pm$ 0.4169 \\
				\midrule
				MobileNetV2 & glioma & 0.3026 $\pm$ 0.0558 & 0.8517 $\pm$ 0.1229 & 0.4418 $\pm$ 0.0494 & - \\
				& meningioma & 0.6667 $\pm$ 0.5774 & 0.0120 $\pm$ 0.0180 & 0.0233 $\pm$ 0.0348 & - \\
				& no\_tumor & 0.5343 $\pm$ 0.3066 & 0.6548 $\pm$ 0.1750 & 0.5249 $\pm$ 0.0899 & - \\
				& pituitary & 0.1412 $\pm$ 0.2445 & 0.0400 $\pm$ 0.0693 & 0.0623 $\pm$ 0.1080 & - \\
				& Macro Avg & 0.4112 $\pm$ 0.2078 & 0.3896 $\pm$ 0.0345 & 0.2631 $\pm$ 0.0307 & - \\
				& Weighted Avg & 0.3980 $\pm$ 0.2379 & 0.3237 $\pm$ 0.0264 & 0.3237 $\pm$ 0.0264 & 0.2115 $\pm$ 0.0366 \\
				\midrule
				MobileNetV3 & glioma & 0.8912 $\pm$ 0.0353 & 0.9777 $\pm$ 0.0082 & 0.9321 $\pm$ 0.0154 & - \\
				& meningioma & 0.9755 $\pm$ 0.0050 & 0.8639 $\pm$ 0.0334 & 0.9160 $\pm$ 0.0175 & - \\
				& no\_tumor & 0.9445 $\pm$ 0.0308 & 1.0000 $\pm$ 0.0000 & 0.9713 $\pm$ 0.0165 & - \\
				& pituitary & 0.9679 $\pm$ 0.0073 & 0.9733 $\pm$ 0.0208 & 0.9706 $\pm$ 0.0138 & - \\
				& Macro Avg & 0.9448 $\pm$ 0.0148 & 0.9537 $\pm$ 0.0113 & 0.9475 $\pm$ 0.0140 & - \\
				& Weighted Avg & 0.9475 $\pm$ 0.0123 & 0.9447 $\pm$ 0.0140 & 0.9447 $\pm$ 0.0140 & 0.9442 $\pm$ 0.0142\\
				\midrule
				EfficientNetB0 & glioma & 0.9960 $\pm$ 0.0000 & 0.9882 $\pm$ 0.0000 & 0.9921 $\pm$ 0.0000 & - \\
				& meningioma & 0.9934 $\pm$ 0.0000 & 0.9869 $\pm$ 0.0000 & 0.9902 $\pm$ 0.0000 & - \\
				& no\_tumor & 0.9929 $\pm$ 0.0000 & 1.0000 $\pm$ 0.0000 & 0.9964 $\pm$ 0.0000 & - \\
				& pituitary & 0.9868 $\pm$ 0.0000 & 0.9967 $\pm$ 0.0000 & 0.9917 $\pm$ 0.0000 & - \\
				& Macro Avg & 0.9923 $\pm$ 0.0000 & 0.9929 $\pm$ 0.0000 & 0.9926 $\pm$ 0.0000 & - \\
				& Weighted Avg & 0.9920 $\pm$ 0.0000 & 0.9920 $\pm$ 0.0000 & 0.9920 $\pm$ 0.0000 & 0.9920 $\pm$ 0.0000 \\
				\midrule
				EfficientNetB1 & glioma & 0.9987 $\pm$ 0.0023 & 0.9921 $\pm$ 0.0000 & 0.9954 $\pm$ 0.0011 & - \\
				& meningioma & 0.9933 $\pm$ 0.0001 & 0.9750 $\pm$ 0.0136 & 0.9840 $\pm$ 0.0070 & - \\
				& no\_tumor & 0.9976 $\pm$ 0.0041 & 1.0000 $\pm$ 0.0000 & 0.9988 $\pm$ 0.0021 & - \\
				& pituitary & 0.9773 $\pm$ 0.0116 & 1.0000 $\pm$ 0.0000 & 0.9885 $\pm$ 0.0059 & - \\
				& Macro Avg & 0.9918 $\pm$ 0.0036 & 0.9918 $\pm$ 0.0034 & 0.9917 $\pm$ 0.0036 & - \\
				& Weighted Avg & 0.9905 $\pm$ 0.0040 & 0.9903 $\pm$ 0.0042 & 0.9903 $\pm$ 0.0042 & 0.9903 $\pm$ 0.0042 \\
				\midrule
				EfficientNetB2 & glioma & 0.9919 $\pm$ 0.0040 & 0.9712 $\pm$ 0.0164 & 0.9814 $\pm$ 0.0091 & - \\
				& meningioma & 0.9699 $\pm$ 0.0128 & 0.9782 $\pm$ 0.0105 & 0.9740 $\pm$ 0.0084 & - \\
				& no\_tumor & 0.9906 $\pm$ 0.0107 & 1.0000 $\pm$ 0.0000 & 0.9953 $\pm$ 0.0054 & - \\
				& pituitary & 0.9879 $\pm$ 0.0082 & 0.9922 $\pm$ 0.0077 & 0.9900 $\pm$ 0.0017 & - \\
				& Macro Avg & 0.9851 $\pm$ 0.0054 & 0.9854 $\pm$ 0.0047 & 0.9852 $\pm$ 0.0051 & - \\
				& Weighted Avg & 0.9838 $\pm$ 0.0049 & 0.9837 $\pm$ 0.0049 & 0.9837 $\pm$ 0.0049 & 0.9837 $\pm$ 0.0049\\
				\midrule
				Xception & glioma & 0.0847 $\pm$ 0.1466 & 0.3333 $\pm$ 0.5774 & 0.1350 $\pm$ 0.2339 & - \\
				& meningioma & 0.0000 $\pm$ 0.0000 & 0.0000 $\pm$ 0.0000 & 0.0000 $\pm$ 0.0000 & - \\
				& no\_tumor & 0.0933 $\pm$ 0.0808 & 0.6667 $\pm$ 0.5774 & 0.1637 $\pm$ 0.1418 & - \\
				& pituitary & 0.0000 $\pm$ 0.0000 & 0.0000 $\pm$ 0.0000 & 0.0000 $\pm$ 0.0000 & - \\
				& Macro Avg & 0.0445 $\pm$ 0.0165 & 0.2500 $\pm$ 0.0000 & 0.0747 $\pm$ 0.0230 & - \\
				& Weighted Avg & 0.0346 $\pm$ 0.0259 & 0.1780 $\pm$ 0.0658 & 0.1780 $\pm$ 0.0658 & 0.0572 $\pm$ 0.0395 \\
				\midrule
				VGG16 & glioma & 0.9803 $\pm$ 0.0150 & 0.9396 $\pm$ 0.0741 & 0.9582 $\pm$ 0.0335 & - \\
				& meningioma & 0.9427 $\pm$ 0.0509 & 0.9684 $\pm$ 0.0019 & 0.9549 $\pm$ 0.0257 & - \\
				& no\_tumor & 0.9790 $\pm$ 0.0068 & 0.9976 $\pm$ 0.0041 & 0.9882 $\pm$ 0.0020 & - \\
				& pituitary & 0.9867 $\pm$ 0.0099 & 0.9822 $\pm$ 0.0117 & 0.9844 $\pm$ 0.0051 & - \\
				& Macro Avg & 0.9722 $\pm$ 0.0132 & 0.9720 $\pm$ 0.0172 & 0.9714 $\pm$ 0.0162 & - \\
				& Weighted Avg & 0.9706 $\pm$ 0.0157 & 0.9693 $\pm$ 0.0177 & 0.9693 $\pm$ 0.0177 & 0.9692 $\pm$ 0.0178 \\
				\midrule
				VGG19 & glioma & 0.9484 $\pm$ 0.0351 & 0.9541 $\pm$ 0.0421 & 0.9502 $\pm$ 0.0081 & - \\
				& meningioma & 0.9624 $\pm$ 0.0130 & 0.9434 $\pm$ 0.0068 & 0.9527 $\pm$ 0.0030 & - \\
				& no\_tumor & 0.9725 $\pm$ 0.0231 & 0.9976 $\pm$ 0.0041 & 0.9848 $\pm$ 0.0111 & -\\
				& pituitary & 0.9744 $\pm$ 0.0287 & 0.9745 $\pm$ 0.0267 & 0.9739 $\pm$ 0.0039 & -\\
				& Macro Avg & 0.9645 $\pm$ 0.0061 & 0.9674 $\pm$ 0.0041 & 0.9654 $\pm$ 0.0047 & - \\
				& Weighted Avg & 0.9639 $\pm$ 0.0039 & 0.9630 $\pm$ 0.0044 & 0.9630 $\pm$ 0.0044 & 0.9629 $\pm$ 0.0044 \\
				\midrule
				InceptionV3 & glioma & 0.6564 $\pm$ 0.5686 & 0.5315 $\pm$ 0.4983 & 0.5778 $\pm$ 0.5128 & - \\
				& meningioma & 0.8972 $\pm$ 0.1694 & 0.6645 $\pm$ 0.5132 & 0.6401 $\pm$ 0.4458 & - \\
				& no\_tumor & 0.5887 $\pm$ 0.4180 & 0.9905 $\pm$ 0.0165 & 0.6719 $\pm$ 0.3794 & -\\
				& pituitary & 0.6629 $\pm$ 0.5741 & 0.5722 $\pm$ 0.5136 & 0.6104 $\pm$ 0.5343 & -\\
				& Macro Avg & 0.7013 $\pm$ 0.3670 & 0.6897 $\pm$ 0.3750 & 0.6250 $\pm$ 0.4676 & -\\
				& Weighted Avg & 0.7225 $\pm$ 0.3490 & 0.6487 $\pm$ 0.4312 & 0.6487 $\pm$ 0.4312 & 0.6198 $\pm$ 0.4797\\
				
				\bottomrule
			\end{tabular}
		}
	}
	\label{tab:class}
\end{table}

\subsubsection*{Classification Results}
\color{black}
To evaluate the classification performance on our newly introduced brain tumor dataset, we conducted a comprehensive analysis of several baseline models for classifying brain tumor types: \textit{Glioma}, \textit{Meningioma}, \textit{Pituitary}, and \textit{non-tumorous}. The evaluated models include ResNet50, ResNet101, DenseNet121, DenseNet169, MobileNetV2, MobileNetV3, EfficientNetB0, EfficientNetB1, EfficientNetB2, Xception, VGG16, VGG19, and InceptionV3. Each model was trained and tested three times to ensure robust and reliable results, with performance reported as the mean and standard deviation of key metrics: Precision, Recall, F1-Score, and Accuracy.

The evaluation metrics were computed per class, alongside macro and weighted averages, to provide a comprehensive view of model performance across diverse tumor types. The macro average treats all classes equally, while the weighted average accounts for class imbalance by weighting each class’s contribution based on the number of samples, offering a realistic assessment of performance in clinical scenarios where tumor type distributions may vary.

As presented in Table~\ref{tab:class}, the benchmarks show that high classification accuracy is achievable on this dataset. EfficientNetB0 performs strongly, with a weighted average F1-score of 0.9920 $\pm$ 0.0000 and an accuracy of 0.9920 $\pm$ 0.0000, achieving perfect recall (1.0000 $\pm$ 0.0000) for the \textit{non-tumorous} class. EfficientNetB1 follows closely with a weighted F1-score of 0.9903 $\pm$ 0.0042, while ResNet50 and MobileNetV3 deliver competitive results (weighted F1-scores of 0.9820 $\pm$ 0.0080 and 0.9447 $\pm$ 0.0140, respectively). 

In contrast, Xception exhibits the lowest performance, with a weighted F1-score of 0.1780 $\pm$ 0.0658, failing entirely on meningioma and pituitary (F1-score: 0.0000 $\pm$ 0.0000). Similarly, DenseNet121 and DenseNet169 show unstable performance, with high standard deviations, indicating limited generalizability. MobileNetV2 also struggles, particularly with meningioma (recall: 0.0120 $\pm$ 0.0180), likely due to insufficient model capacity.

The VGG variants (VGG16 and VGG19) achieve moderate performance, with weighted F1-scores of 0.9693 $\pm$ 0.0177 and 0.9630 $\pm$ 0.0044, respectively, while InceptionV3 shows inconsistent results (weighted F1-score: 0.6487 $\pm$ 0.4312), reflecting challenges in handling complex tumor morphology or class imbalances.

This evaluation underscores the strong performance of EfficientNet models, particularly EfficientNetB0, which combines high accuracy with remarkable stability across all tumor types. The results validate the utility of our dataset for developing reliable diagnostic tools, while the stark performance differences across architectures emphasize the importance of model selection in medical imaging tasks, where precision and consistency are critical. This work establishes a robust benchmark for brain tumor classification and provides a foundation for future research to explore diverse models and training protocols using this dataset.

\section*{Data Availability}

We introduce a new dataset, BRISC (Brain Tumor MRI Dataset for Segmentation and Classification), which is publicly available at Kaggle \url{https://www.kaggle.com/datasets/briscdataset/brisc2025/}.

\color{black}

\section*{Code Availability}

The custom code developed for the baseline models are publicly available at the BRISC dataset repository on Kaggle \url{https://www.kaggle.com/datasets/briscdataset/brisc2025/}.

\section*{Acknowledgements}

We thank Fatemeh Gheisari, our radiologist, for her invaluable assistance with expert annotations and guidance throughout this study.

\section*{Author Contributions}
Amirreza Fateh wrote the original draft of the manuscript, supervised the research, and contributed to the methodology. Yasin Rezvani contributed to the dataset collection, curation, and methodology, and implemented the code. Sara Moayedi contributed to the dataset creation. Sadjad Rezvani contributed to the methodology and implementing the baselines. Fatemeh Fateh, as the consulting physician, supervised the dataset design and labeling and contributed to writing the manuscript. Mansoor Fateh supervised the research, reviewed the manuscript, and contributed to the methodology. Vahid Abolghasemi reviewed the manuscript and supervised the work. All authors reviewed and approved the final manuscript.

\section*{Competing Interests}

The authors declare no competing interests.

\bibliographystyle{IEEEtran}

\end{document}